\documentclass[aip, rsi, reprint, altaffilletter, floatfix]{revtex4-1}
\usepackage{graphicx}
\usepackage{float}
\usepackage{amsmath}
\usepackage{color}
\usepackage{multirow}

\begin{document}


\title{
Parallax diagnostics of radiation source geometric dilution for iron opacity experiments
}
\thanks{Contributed paper published as part of the Proceedings of the 20th Topical Conference on High-Temperature Plasma Diagnostics, Atlanta, Georgia, June, 2014.\\}



\author{T. Nagayama}
\affiliation{Sandia National Laboratories, Albuquerque, New Mexco 87185, USA}

\author{J.E. Bailey}
\affiliation{Sandia National Laboratories, Albuquerque, New Mexco 87185, USA}

\author{G. Loisel}
\affiliation{Sandia National Laboratories, Albuquerque, New Mexco 87185, USA}

\author{G.A. Rochau}
\affiliation{Sandia National Laboratories, Albuquerque, New Mexco 87185, USA}

\author{R.E. Falcon}
\affiliation{Sandia National Laboratories, Albuquerque, New Mexco 87185, USA}


\date{\today}

\begin{abstract}
Experimental tests are in progress to evaluate the accuracy of the modeled iron opacity at solar interior conditions [J.E. Bailey et al., Phys. Plasmas 16, 058101 (2009)]. The iron sample is placed on top of the Sandia National Laboratories z-pinch dynamic hohlraum (ZPDH) radiation source. The samples are heated to 150 -  200 eV electron temperatures and 7$\mathrm{\times 10^{21}}$ - 4$\mathrm{\times 10^{22}}$ cm$^{-3}$ electron densities by the ZPDH radiation and backlit at its stagnation [T. Nagayama et al., Phys. Plasmas 21, 056502 (2014)]. The backlighter attenuated by the heated sample plasma is measured by four spectrometers along $\pm$ 9$^\circ$ with respect to the z-pinch axis to infer the sample iron opacity. Here we describe measurements of the source-to-sample distance that exploit the parallax of spectrometers that view the half-moon-shaped sample from $\pm$ 9$^\circ$. The measured sample temperature decreases with increased source-to-sample distance. This distance must be taken into account for understanding the sample heating.
\end{abstract}


\maketitle 

\section{Introduction}

Opacity quantifies photon absorption in matter and plays a crucial
role in many high energy density plasmas, including inertial fusion
plasmas and stellar interiors\cite{mihalas1978stellar}. Modeling
opacities of ions with multiple bound electrons is very challenging
and employs approximations that need to be experimentally validated\cite{Perry:1996hy,Bailey:2009hh}.
Performing reliable opacity experiments is also challenging and must
satisfy many criteria\cite{Perry:1996hy,Bailey:2009hh}. Measuring
opacity becomes more difficult at higher temperature because the opacity
sample has to be heated to the high temperature without significant
gradients and has to be backlit by a bright radiation to minimize
the effect of the hot sample plasma emission on the absorption measurement.
The Sandia National Laboratories (SNL) Z machine (Z) provides a unique
platform to perform opacity experiments at temperatures above 150
eV\cite{Bailey:2007fq}. 

The Z-pinch dynamic hohlraum (ZPDH) is a terawatt x-ray radiation
source at Z that makes high-temperature opacity measurements possible\cite{Rochau:2014ha}.
The opacity sample is located above the ZPDH radiation source and
is radiatively heated. Most of the photons have energies above 600
eV. This powerful radiation streams through the sample and heats it
without significant gradients\cite{Nagayama:2014kl}. The ZPDH also
provides a bright backlighter to mitigate the sample self-emission.
Recently, we found that the opacity sample can reach higher temperatures
and densities using the same radiation source only by changing the
target configuration\cite{Nash:2010cn,Nagayama:2014kl}. However,
it was not clear why the change in the target configuration affects
the sample temperature if the sample is heated by the same radiation
source. To further optimize this high temperature opacity experimental
platform, it is crucial to understand what dictates the sample temperature.
In this article, we provide experimental evidence that the source-to-sample
distance depends on the sample configuration. This distance controls
the source geometric dilution at the sample, thereby affecting the
sample temperature. 

\begin{figure}
\includegraphics[width=8cm]{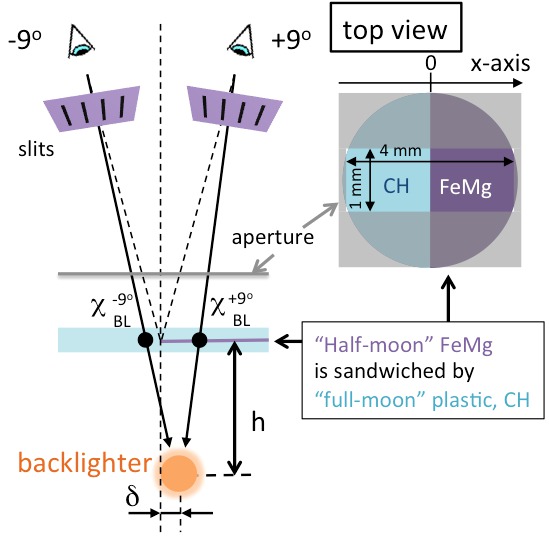}

\caption{(Color online) Two space-resolving spectrometers located at $\pm9^{\circ}$with
respect to the z-axis record the sample-transmitted backlighter images.
Due to the angle difference, the spectrometers at $\pm$9$^{\circ}$
see the backlighter centered at different locations on the sample
(i.e., $x_{BL}^{+9^{\circ}}$ and $x_{BL}^{-9^{\circ}}$). This parallax
not only measures FeMg-attenuated and -unattenuated spectra simultaneously,
but it also characterizes the backlighter relative location with respect
to the ``half-moon'' boundary, $h$ and $\delta$. }
\end{figure}

\section{SNL Opacity experiments and parallax}

The typical SNL opacity experimental setup is shown in Fig. 1. The
target consists of a semi-circular FeMg sample sandwiched by a circular
tamping material (e.g., plastic, CH), which we call a \textquotedblleft{}half-moon\textquotedblright{}
target. Mg is mixed in the Fe sample to diagnose the Fe conditions
(i.e., electron temperature, $T_{e}$, and electron density, $n_{e}$)
using Mg K-shell spectroscopy\cite{Bailey:2008cb,Nagayama:2014kl}.
This target is placed above the ZPDH radiation source, and the ZPDH
radiation heats and backlights the sample\cite{Rochau:2014ha,Bailey:2009hh}.
The backlighter attenuated through the target is recorded by potassium
acid phthalate crystal (KAP) spectrometers fielded along $\pm9^{\circ}$
from the z-axis\cite{Bailey:2008cb}. An aperture above the target
limits the spectrometers\textquoteright{} views to a 4 mm $\times$
1 mm area. Each spectrometer has 4 - 6 slits, each 50 $\mu m$ in
width, at the halfway distance to the sample to provide spatial resolution
of $\sim$0.1 mm along the aperture direction with a magnification
of $\sim$1. The transmitted backlighter images are recorded on Kodak
2492 x-ray films with spatial and spectral resolution. 

Due to the finite source-to-sample distance, $h$, the spectrometer
at $+9^{\circ}$ observes the backlighter bright spot through the
FeMg embedded side at $x_{BL}^{+9^{\circ}}$, while the one at $-9^{\circ}$
observes it on the CH-only side at $x_{BL}^{-9^{\circ}}$ (black dots
in Fig. 1). This spectrometer configuration measures the FeMg-attenuated
and -unattenuated spectra simultaneously, providing FeMg transmission
spectra in a single experiment (shot). However, taking advantage of
this parallax, we can also infer the backlighter location with respect
to the ``half-moon'' boundary (i.e., $h$ and $\delta$ in Fig.
1) based on $x_{BL}^{+9^{\circ}}$ and $x_{BL}^{-9^{\circ}}$ as follows:
\begin{equation}
h=\frac{x_{BL}^{+9^{\circ}}-x_{BL}^{-9^{\circ}}}{2\mathrm{tan}(9^{\circ})},\ \delta=\frac{1}{2}\left(x_{BL}^{+9^{\circ}}+x_{BL}^{-9^{\circ}}\right)\label{eq:h}
\end{equation}
assuming that the source-to-detector distance is much larger than
$x_{BL}^{+9^{\circ}}$ - $x_{BL}^{-9^{\circ}}$. 

To extract $x_{BL}^{+9^{\circ}}$ and $x_{BL}^{-9^{\circ}}$ from
the data, one has to understand the emergent intensity spatial profiles
measured at $\pm9^{\circ}$ (Fig. 2). The x-axis is defined such that
the ``half-moon'' boundary is at x=0 and the FeMg-embedded region
is at x$>$0. The hypothetical transmission spatial profile at a given
wavelength (blue) is systematically lower at x$>$0 due to the FeMg
attenuation. The apparent backlighter spatial profiles (green) are
centered at different locations with respect to the sample for the
$\pm9^{\circ}$ spectrometers (i.e., $x_{BL}^{+9^{\circ}}$ and $x_{BL}^{-9^{\circ}}$,
respectively). While most of the backlighter spatial profile is attenuated
through the FeMg region on the $+9^{\circ}$ spectrometer, only the
backlighter wing is attenuated through FeMg on the $-9^{\circ}$ spectrometer.
As a result, one expects to see a double peak in the emergent spatial
profile at $+9^{\circ}$, while one expects a skewed single peak at
$-9^{\circ}$. 

\begin{figure}
\includegraphics[width=8cm]{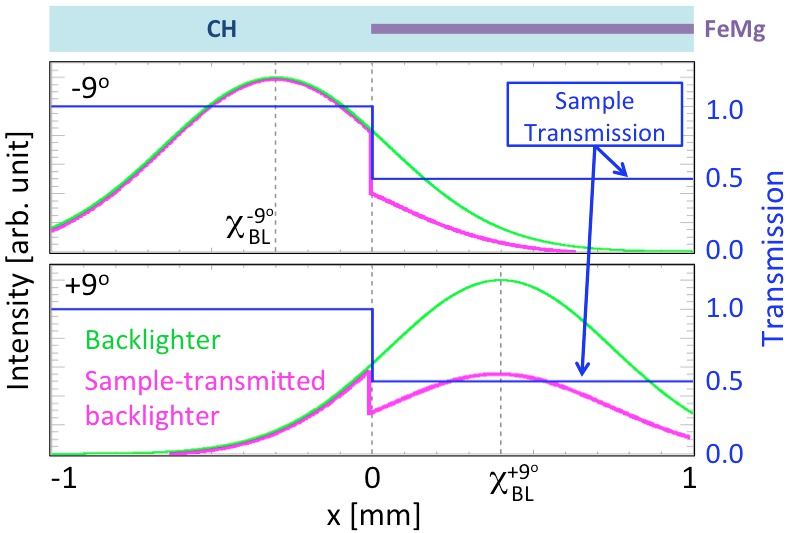}

\caption{(Color online) Idealized schematics illustrate how the backlighter
(green) observed at different angles results in different emergent
intensity spatial profiles (magenta). }

\end{figure}

Figure 3 shows the data recorded by the spectrometers at $\pm9^{\circ}$.
Each image is the average over four slit images to improve the signal-to-noise
ratio and to average out random defects in the individual slit images\cite{Nagayama:2014kl}.
The horizontal (spectral) and vertical (spatial) axes are produced
by the KAP crystals and the slits, respectively. The dark vertical
lines correspond to Fe or Mg bound-bound absorption lines. The image
recorded at $+9^{\circ}$ shows longer Fe and Mg lines than those
recorded at $-9^{\circ}$ due to the apparent backlighter peak locations
(Fig. 2). 

In order to measure $x_{BL}^{+9^{\circ}}$ and $x_{BL}^{-9^{\circ}}$,
one has to extract the locations of the \textquotedblleft{}half-moon\textquotedblright{}
boundary and the apparent backlighter peak. To objectively extract
them, we take a spatial lineout on a strong bound-bound absorption
line. The magenta curves in Fig. 4 show an example for the Mg He$\alpha$
line (i.e., absorption due to $1s^{2}-1s2p$ He-like Mg transition)
at $\sim$9.17$\mathrm{\AA}$ (lineout $\Delta\lambda$=0.02 $\mathrm{\AA}$).
As discussed earlier, the magenta curve at $+9^{\circ}$ has a double
peak, while the one at $-9^{\circ}$ has a skewed single peak. We
approximate the spatial profile in the absence of the Mg He$\alpha$
(green curves in Fig. 3) by averaging two spatial lineouts taken on
each side of the Mg He$\alpha$ line. The lineout locations for the
$+9^{\circ}$ image are indicated by vertical green dashed lines in
Fig. 3. The He$\alpha$ line transmission is determined from the ratio
of the magenta and green curves. The resultant transmission spatial
profiles clearly show low-transmission FeMg embedded regions, and
the x-axis is defined from its inflection point. We note that the
``half-moon'' boundary is not as sharp as the one in Fig. 2. This
is because of the instrument spatial resolution and the sample hydrodynamics
integrated over the backlighter duration. Once $x_{BL}^{+9^{\circ}}$
and $x_{BL}^{-9^{\circ}}$ are defined by the apparent backlighter
peak locations on the defined x-axis, the backlighter location, $h$
and $\delta$, can be estimated from Eq. \ref{eq:h}.

\begin{figure}
\includegraphics[width=8cm]{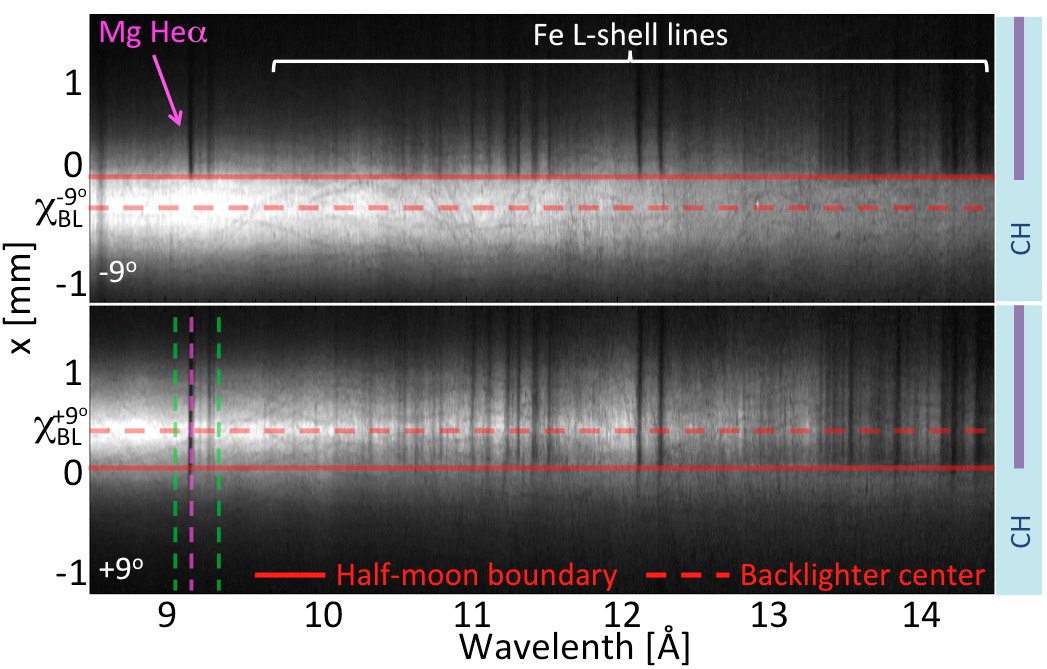}

\caption{(Color online) Backlighter images attenuated through the target recorded
from $-9^{\circ}$ and $+9^{\circ}$. FeMg is embedded at x$>$0.
Horizontal red solid and dashed lines indicate the locations of the
\textquotedblleft{}half-moon\textquotedblright{} boundary at x=0 and
the apparent backlighter peaks at x=$x_{BL}^{-9^{\circ}}$ and x=$x_{BL}^{+9^{\circ}}$,
respectively. }
\end{figure}

\begin{figure}
\includegraphics[width=8cm]{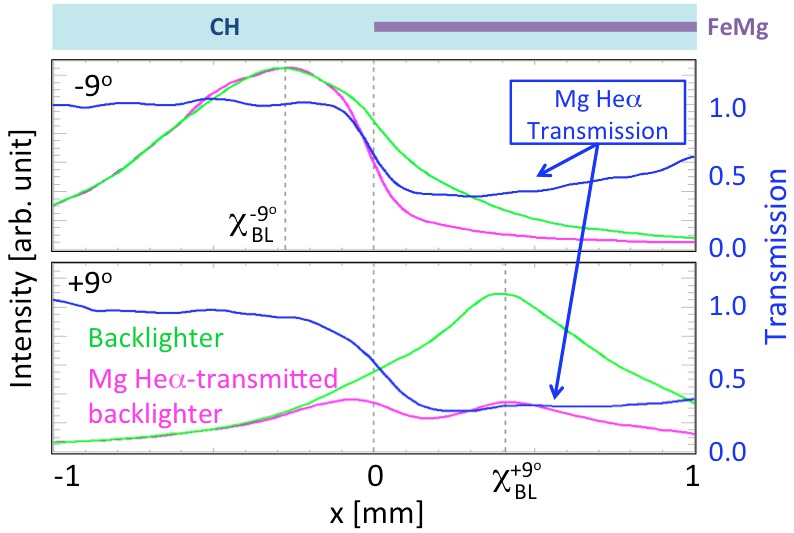}

\caption{(Color online) The spatial lineouts are extracted from FIG. 3 at Mg
He$\alpha$ (magenta) and its nearby continuum (green). Mg He$\alpha$
bound-bound line transmission spatial lineouts (blue) are extracted
by dividing the magenta by the green. The ``half-moon'' boundary
and the x-axis are defined based on the transmission spatial lineouts,
and $x_{BL}^{-9^{\circ}}$ and $x_{BL}^{+9^{\circ}}$ are defined
based on the continuum peaks. }

\end{figure}

\section{Results}

Parallax is systematically applied to ten Fe opacity shots performed
under different sample configurations \cite{Nagayama:2014kl}. There
are three different CH configurations and multiple different Fe thicknesses
for each configuration. There is one shot where the sample is raised
by 1.5 mm from its nominal location. For each shot, parallax is applied
to the available bound-bound lines that are strong enough to define
the ``half-moon'' boundary from their line-transmission spatial
profiles. The number of usable lines depends on their areal density,
Stark line width, and the spectral range of the spectrometers used.
For each shot, the mean h and its standard deviation are computed
from parallax results of two to seven Fe and Mg lines. Parallax results
from Fe and Mg lines agree with each other. The validity of this uncertainty
estimate was also verified from the shots with four spectrometers,
two each at $+9^{\circ}$ and $-9^{\circ}$. Figure 5 summarizes the
measured $h$ as a function of $T_{e}$ inferred from Mg K-shell spectroscopy\cite{Nagayama:2014kl}.
We confirm a strong anti-correlation between $h$ and $T_{e}$ (Pearson
correlation coefficient = -0.91). 

To investigate this point synthetically, we use a 3D view factor code
VISRAD\cite{MacFarlane:2003jc} and a calibrated ZPDH intensity image
from one of our experiments to calculate the heating radiation at
the sample as a function of the sample distance from the ZPDH radiation
source. The details of the calculation will be discussed elsewhere.
The blue curve in Fig. 5 shows the resultant radiation brightness
temperature, $T_{B}$, as a function of $h$. This result suggests
that the radiation source heats the sample to different temperatures
due to source radiation geometric dilution. The radiation brightness
temperature is systematically higher than $T_{e}$ due to the complex
heating mechanism involving radiation transport and hydrodynamics
and beyond the scope of this article. 

Figure 5 also shows that, for similar $T_{e}$, shot-to-shot variation
in the inferred $h$ is larger than the individual measurement uncertainties
due to the 3-D radiation transport effects of the backlighter. While
Eq. (1) is derived assuming an instantaneous point backlighter, the
actual backlighter emission is a result of the radiation transport
through the 3-D ZPDH plasma, which spatially varies over a few ns
duration. Thus, the variation in the inferred $h$ comes from the
irreproducibility in the evolution of 3-D ZPDH plasma and the resultant
irreproducibility in the line-of-sight dependent effects on the measurements.

We found that $h$ was anti-correlated to $T_{e}$ and confirmed that
the sample reached a different temperature due to the geometric dilution
of the radiation source. The parallax results are important i) to
better understand our platform and further optimize SNL Z opacity
experiments and ii) to better understand the sample heating and accurately
evaluate how close our sample is to local thermal equilibrium. 

\begin{figure}[t]
\includegraphics[width=8cm]{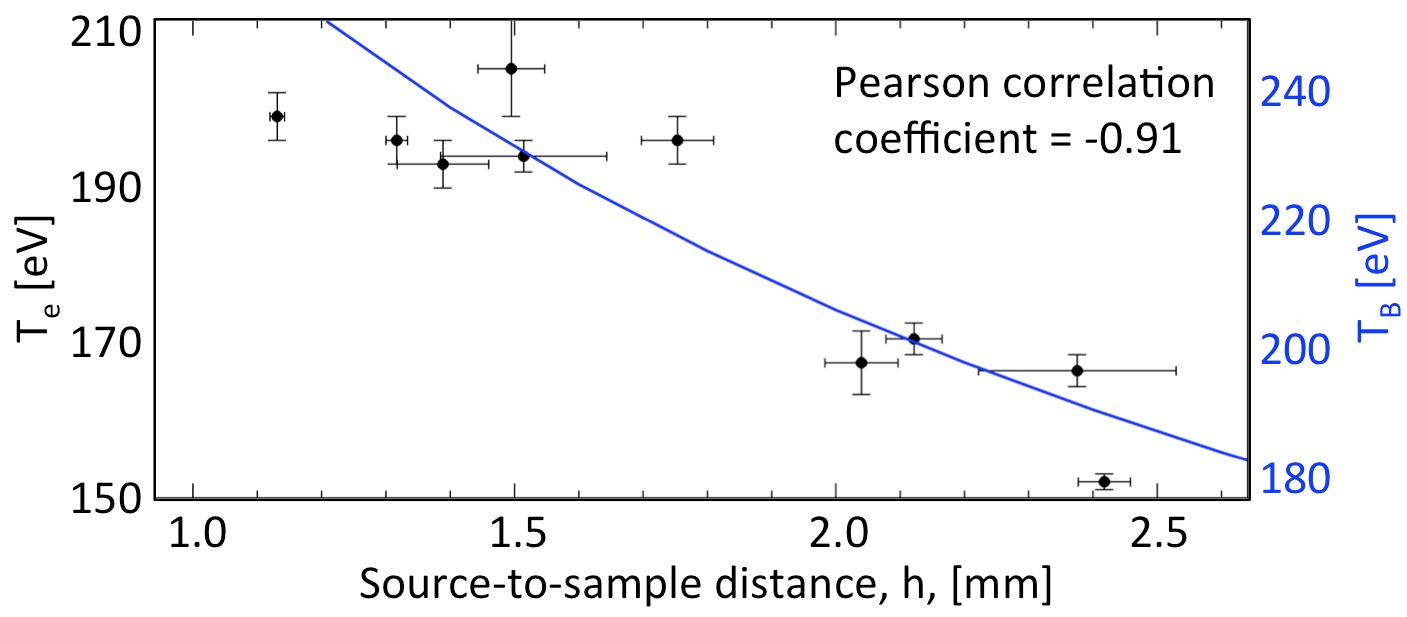}

\caption{(Color online) There is a strong correlation between the measured
electron temperature, $T_{e}$, and the measured source-to-sample
distance, $h$. The blue curve is a modeled radiation brightness temperature
as a function of $h$.}
\end{figure}

\section*{ACKNOWLEDGMENTS }

Sandia is a multiprogram laboratory operated by Sandia Corporation,
a Lockheed Martin Company, for the (U.S.) Department of Energy under
Contract No. DE-AC04-94AL85000.

%

\end{document}